\documentclass[aps,prb,reprint,superscriptaddress]{revtex4-1}
\usepackage{graphicx} 
\usepackage{float} 

\begin{document}
\preprint{IGC/MSG/JNL/2011/60}

\title{Dislocation Driven Chromium Precipitation in Fe-9Cr Binary Alloy: A Positron Lifetime Study}

\author{S. Hari Babu}
\author{R. Rajaraman}
\author{G. Amarendra}
\email[Mail to:]{amar@igcar.gov.in}
\homepage[Visit:]{http://msg.igcar.gov.in}
\author{R. Govindaraj}
\affiliation{Materials Science Group, Indira Gandhi Centre for Atomic Research, 
Kalpakkam - 603102, TN, India}

\author{N. P. Lalla}
\affiliation{UGC-DAE CSR, University Campus, Indore - 452001, MP, India}

\author{Arup Dasgupta}
\affiliation{Physical Metallurgy Group, Indira Gandhi Centre for Atomic Research, Kalpakkam - 603102, TN, India}

\author{Gopal Bhalerao}
\affiliation{UGC-DAE CSR, Kokilamedu-603104, TN, India}

\author{C.S. Sundar}
\affiliation{Materials Science Group, Indira Gandhi Centre for Atomic Research, Kalpakkam - 603102, TN, India}

\date{\today}

\begin{abstract}
The influence of initial heat treatment on anomalous Cr precipitation within high temperature solubility region in Fe-9Cr alloy has been investigated using positron lifetime studies. Air-quenched samples with pre-existing dislocations exhibited a distinct annealing stage in positron lifetime between 800 and 1100 K corresponding to Cr-precipitation.  During this stage, Transmission Electron Microscopy showed fine precipitates of average size 4 nm, dispersed throughout the sample and from EDS analysis they are found to be Cr-enriched. The existence of dislocations is found to be responsible for Cr precipitation. 
\end{abstract}



\maketitle

\section{INTRODUCTION}

Ferritic/Martensitic steels are candidate structural materials for future nuclear reactors having aggressive temperatures and irradiation conditions \cite{cook2006}. Swelling resistance is found to be good for these steels at elevated temperatures \cite{klueh2004}.  Owing to initial thermal treatments, the microstructure consists of lath structure with dislocations and fine precipitates, which is found to be responsible for superior mechanical properties \cite{klueh2004}.  In particular, Fe-9\%Cr alloys are of special interest owing to their similar Cr concentration as that of modified 9Cr-1Mo and Eurofer-97 steels. From fundamental point of view, Fe-Cr phase diagram continues to evoke a lot of interest and getting revised with more precise experimental results and ab-initio calculations, as evident from recent critical review \cite{xiong2010}. In particular, both Fe rich and Cr rich boundaries have attracted wide attention in recent times \cite{xiong2010, bonny2011}. Considerable efforts are also being made to obtain reliable interatomic potentials for computational modeling of Fe-Cr system and a recent report compares various potentials used in calculating Fe rich solid solution boundary \cite{bonny2011}.

Earlier resistivity and diffuse neutron scattering studies showed short range ordering in Fe-Cr alloys for Cr content more than 10at\% \cite{mirebeau2010}. Neutron irradiation induced Cr precipitation was observed in Fe-Cr alloys containing $>$ 9\%Cr \cite{heintze2010,matijasevic2008}. Positron annihilation studies have also showed accelerated phase separation upon electron irradiation followed by isochronal annealing below 700 K for wide range of Cr concentrations (5-20 wt\%) \cite{kuramoto1992, druzhkov2011}. Electron microscopy and Energy dispersive X-ray spectroscopy studies showed clustering of Cr with insitu electron irradiation at 873 K for Fe-2.8at\%Cr \cite{ezawa1994}. Helium implantation causing Cr segregation at grain boundaries was also reported \cite{krsjak2008}. Metropolis Monte Carlo simulations and positron measurements with electron irradiated conditions showed signatures for Cr clustering \cite{kwon2009} and it also supported the study showing tendency to resist the formation of a Cr-vacancy complex \cite{olsson2007}. As a whole, the precipitation of Cr depends on the amount of Cr, type of irradiation and irradiation temperature. Except for Ref. \cite{ezawa1994}, which is irradiation at higher temperature (873 K), electron and neutron irradiation showed precipitation for higher Cr alloys ($>$9$\%$) only, while low Cr alloys showed Cr precipitation only with ion irradiation, as it can produce irradiation cascades. Contrary to this, simulation work showed tendency for clustering even for pure thermal aging \cite{bonny2009,terentev2009,lu2010}, which is not observed experimentally upto now. 

Positron annihilation technique is a powerful characterization technique to probe open volume defects such as vacancies, vacancy clusters, dislocations and grain/lath/precipitate boundaries. In particular, positron lifetime, which is inversely proportional to the local electron density at the site of positron annihilation, is quite sensitive to the nature and size of defects. As matrix/precipitate interfaces act as potential trapping site and thus, precipitation occurring at nano-scale can be effectively probed by positron annihilation spectroscopy. Earlier positron lifetime studies of TiC precipitation in plastically deformed D9 steels demonstrated the sensitivity to precipitate-matrix interfaces \cite{rajaraman1994}. Incoherent precipitate interfaces act as good trapping sites than coherent precipitates, as the later do not have considerable open volume defects.  In Fe-Cr system, both Fe, Cr having lattice variation of only 0.04 $\rm{\AA}$, incoherent precipitates were not expected at small cluster sizes. Similarly, for fine coherent precipitates of Cr in Fe-Cr matrix, TEM imaging will be very difficult, as summarised by Bonny et al. \cite{bonny2009}.

We have investigated Fe-9(wt\%)Cr system to understand the role of Cr in inducing microstructural changes in ferritic matrix. The present study involves positron  lifetime and transmission electron microscopy studies to bring out experimental evidence of Cr precipitation upon pure heat treatment, and its dependence on initial microstructure in particular dislocations. 

\section{Experimental Details}

\subsection{Sample preparation}

Fe-9(wt\%)Cr alloy was prepared using arc melting and homogenized at 1090 K for 144 hr. The samples were cold rolled to make sheets of thickness 300 $\mu$m. The chemical composition (in wt\%) was found to be Cr (9.50 $\pm$ 0.23) and balance Fe using EPMA analysis. No additional impurities were seen within detectable limit of EPMA ($\sim$ 0.01wt\%). Owing to the importance of C content on the formation of vacancy-carbon complexes \cite{hautojarvi1980} and carbides, carbon content was measured using combustion method and was found to be 65 $\pm$ 10 ppm, which is well below the solubility limit of $\sim$ 450 ppm at room temperature. Samples of 8 $\times$ 12 $\times$ 0.3 mm dimensions were homogenized at 1423 K (inside $\gamma$ loop) for 1 hr followed by air quenching. These samples are termed as AQ (Air-Quenched). One set of homogenized samples were heat treated at 1073 K for 2 hr, which is just below the $\gamma$ boundary and will be referred as HT (Heat Treated) samples. Similarly another set of pure Fe (99.99\%) samples were annealed at 923 K  for 3 hr followed by 20 \% cold working (positron lifetime saturated by 15 \% cold working) and these are termed CW (Cold Worked). All these  samples were subjected to isochronal annealing treatment from 300 to 1323 K in steps of 50 K for 1hr in a vacuum of $<10^{-6}$ Torr, followed by air cooling outside the furnace. Results of CW Fe were reported elsewhere \cite{haribabu2010} and those results are used here to compare with Fe-9Cr alloy. 

\subsection{Positron lifetime measurements}

Positron lifetime measurements were carried out at room temperature with a fast-fast lifetime spectrometer having a time resolution of 260 ps (FWHM) and using $^{22}$Na source. Source lifetime and intensity were measured for subsequent corrections using annealed Fe sample as reference. For each measurement $> 10^6$ counts were accumulated and measured lifetime spectra were analyzed using LT program \cite{kansy1996}. In principle, experimental lifetime spectrum is a linear combination of n exponentials, corresponding to n distinct annihilation sites, with coefficients proportional to fraction of positron annihilating in that particular site. When number of distinct annihilation sites are more than three or the lifetime variation for distinct annihilation sites is relatively less, the statistical fluctuations dominates and lifetime cannot be resolved properly. Mostly the annihilation sites at grain/lath boundaries and dislocations satisfy the above conditions. We had tried fitting the experimental lifetime spectra in terms of two and three lifetime components. While the variance was lower for these fits, the lifetime values and intensities were not physically meaningful. Thus, we have resorted to only single lifetime over entire annealing temperature range in all the measurements.

\subsection{Methodology of theoretical calculations}

Theoretical positron lifetime calculations for various defect and bulk configurations related to present study were carried out using atomic superposition method, as implemented in MIKA-doppler code \cite{doppler}. For enhancement factor of the electron density at the positron, the relation proposed by Boronski and Nieminen \cite{bn1986} was used. All these calculations were carried out in un-relaxed structures. In the case vacancy defect configurations in metals, atoms around vacancies relax inward \cite{makkonen2006}. However, positron induced forces from trapped state at metal vacancies on neighbouring atoms are known to nearly cancel the inward relaxation effects \cite{makkonen2006} and hence calculations of un-relaxed vacancies in present studies should be representatives of experimental conditions.

\subsection{Transmission electron microscopy}

TEM studies were done on AQ samples in as prepared condition and samples subsequently annealed till temperatures of 723, 900, 973 and 1073 K. These samples had undergone identical cumulative annealing treatment as in the case of positron study. TEM studies were carried out, using a Tecnai-G$^2$-20 TEM facility operating at 200 kV. The specimens for TEM study were prepared by mechanical grinding till 100 $\mu$m followed by dimpling till 30 $\mu$m. Final thinning of the sample was done using Ar ion-beam polishing at 3-5 kV/20-40 $\mu$A and at a grazing incidence of 3$^o$ with respect to the sample surface. Analytical Transmission Electron Microscopy (ATEM), Philips CM 200 microscope, with operating voltage of 200 kV was used for obtaining Selected Area Diffraction (SAD) and chemical characterization of precipitates using Energy-dispersive X-ray spectroscopy (EDS) on Fe-Cr sample aged at 973 K.

\section{results and discussions}

Fig-1 shows (a) mean positron lifetime results of Fe-9Cr samples (AQ and HT) and (b) cold worked pure Fe (CW) as a function of isochronal annealing temperature. As the initial heat treatments are different for AQ and HT samples, they show distinct positron lifetime variations till 1073 K. 
\begin{figure}
\begin{center}
\includegraphics*[width=7.5cm]{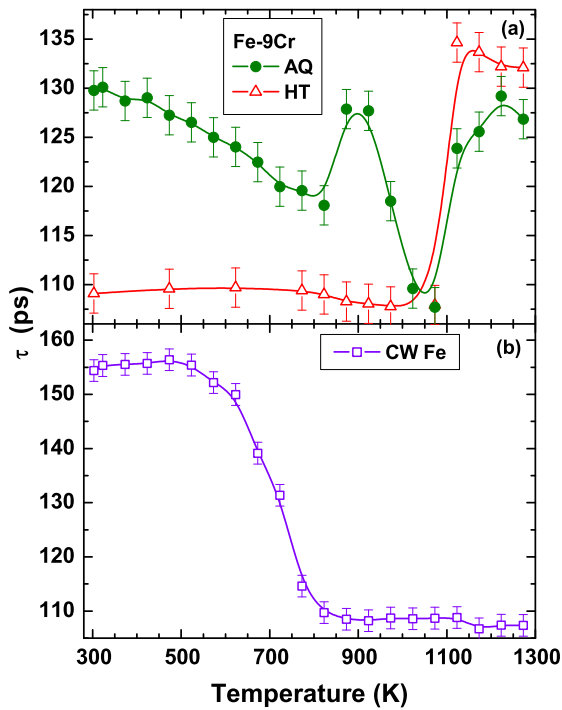}
\end{center}
\caption{\label{fig1} Variation of mean positron lifetime in (a) Fe-9Cr alloys (AQ-filled circles and HT-open triangles) and (b) Cold worked Pure Fe \cite{haribabu2010} (open squares) as a function of annealing temperature.  The line through the data points is a guide to the eye.}
\end{figure}
We first discuss the results of cold worked pure Fe sample. Cold worked Fe (CW) shows saturation lifetime of 154 ps as seen in Fig-1 (b). Nature of positron trapping in dislocation containing samples is a subject still evoking continued interest \cite{hautojarvi1976,smedskjaer1980,sato2007,petrov2010}. We have calculated positron lifetime for various defect configurations relevant to the present study and Table I lists our computed lifetimes along with other computed and experimental lifetimes available from earlier reports \cite{sato2007,petrov2010,robles2007,hidalgo1992,kamimura1995}. More details of computed positron lifetimes at the dislocation and Fe-Cr interfacial configurations will be discussed elsewhere\cite{haribabu2011}. It is clear from the Table I that pure dislocation can not explain the observed positron lifetime in cold worked Fe. On the other hand, lifetime value of vacancies trapped at dislocation core compares well with our experimental value and hence, the saturation positron lifetime observed in cold worked Fe is assigned to vacancies trapped in dislocations. Henceforth, it is implied that positron trapping or annihilation at dislocation means annihilation at vacancies associated with  dislocations. Present lifetime value is in consistent with reported experimental value of 158 ps for deformed Fe \cite{Krause-Rehberg2009}. This lifetime remains constant till 523 K and starts decreasing sharply beyond 673 K. The decrease in lifetime between 523  and 823 K can be understood as annealing of dislocations along with associated vacancies. It reaches defect-free Fe value, i.e., 108 ps, beyond 823 K indicating complete recrystallization.  Above 1185 K iron will get converted to FCC phase. Though there is structural phase transition from FCC to BCC in the cooling process, when the heat treatment temperature is beyond 1185 K, lifetime shows only marginal change from bulk value of BCC phase \cite{haribabu2010}. 

\begin{table*}
\caption{\label{table1} Calculated positron lifetimes for various defect configurations in FeCr alloy.}
\begin{ruledtabular}
\begin{tabular}{lccc}
System & Computed Lifetime (ps) & 
Computed LT (ps) & Experimental LT (ps)  \\
 & (this work) & (previous work) & (previous work) \\
\hline
Perfect Fe & 
104 & 
108\footnotemark[1] & 
111\footnotemark[2] \\ 
Perfect Cr & 
109 & 
104-118\footnotemark[2] & 
120\footnotemark[2] \\
Vacancy in bulk Fe & 
177 & 
179\footnotemark[1] & 
175\footnotemark[2] \\
Vacancy in bulk Cr & 
178 & 
188\footnotemark[2] & 
150\footnotemark[2] \\
Edge dislocation in Fe & 
121 & 
118\footnotemark[1], 117\footnotemark[3] & 
\\
Jog on the edge dislocation in Fe & 
 & 
117\footnotemark[3], 117\footnotemark[4] & 
\\
Vacancy at edge dislocation in Fe & 
 & 
146\footnotemark[1], 144\footnotemark[4] & 
150\footnotemark[5] \\
Di-vacancy at edge dislocation in Fe & 
 & 
157\footnotemark[1], 153\footnotemark[4] & 
 \\
Perfect [100] oriented interface of Fe-Cr & 
112 & 
 & 
 \\
Fe Vacancy at [100] oriented interface of Fe-Cr & 
186 & 
 & 
 \\
Cr Vacancy at [100] oriented interface of Fe-Cr & 
188 & 
 & 
 \\
\end{tabular}
\end{ruledtabular}
\footnotetext[1]{Reference \cite{petrov2010}}
\footnotetext[2]{Reference \cite{robles2007}}
\footnotetext[3]{Reference \cite{kamimura1995}}
\footnotetext[4]{Reference \cite{sato2007}}
\footnotetext[5]{Reference \cite{hidalgo1992}}

\end{table*}

In as prepared Fe-9Cr AQ sample (Fig-1(a)), lifetime of 130 ps shows quenched-in defects. Presence of chromium is responsible for these defects as they are absent in pure Fe \cite{haribabu2010}. The nature of defects responsible for this will be discussed with observations from transmission electron microscopy. Lifetime in AQ sample decreases gradually till 823 K indicating annealing of quenched defects. Beyond 823 K, lifetime increases sharply, indicating the formation of fresh positron trapping  defects and then decreases above 973 K.  Unlike pure Fe, lifetime exhibits a noticeable increase beyond 1073 K attaining a lifetime value comparable to as-prepared sample. The HT sample (Fig-1(a)) has lifetime of 108 ps in as-prepared condition and does not change till the heat treatment temperature is in FCC phase of Fe-Cr phase diagram i.e., 1073 K and increases beyond this, similar to AQ sample. 

Figure 2 shows TEM images of AQ samples (a) as prepared, annealed at (b) 900 and (c) 973 K states, respectively.  TEM analysis of as-prepared sample showed fine grain structure with average grain size of the order of 0.1 $\mu$m similar to lath structure in ferritic/martensitic steels \cite{klueh2004}. High density of random dislocations are also seen. The grain size is consistent with earlier observations \cite{matijasevic2008,krsjak2008}. Fine grain structure and dislocations are characteristic of fast cooling across $\gamma$ (FCC) to $\alpha$ (BCC) transition at around 1135 K. 

\begin{figure*}
\begin{center}
\includegraphics*[height=4.5cm]{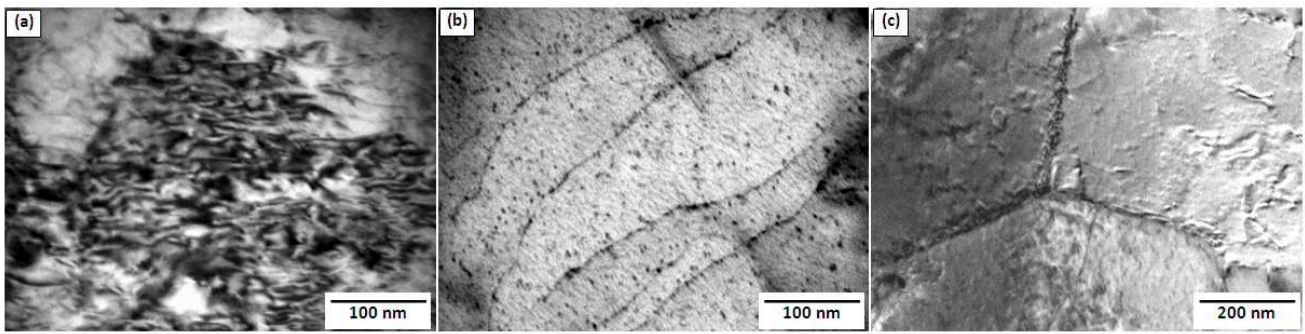}
\end{center}
\caption{\label{fig2} TEM images of (a) air quenched, (b) 900 K  and (c) 973 K  annealed states of AQ samples, respectively.}
\end{figure*}

As the average size of grains is of the order of diffusion length of positrons and they contains dislocations throughout interior of grains as seen in Fig-2a, the higher lifetime of 130 ps in as-prepared sample can be understood arising due to grain boundaries and dislocations. The TEM image taken at 773 K (not included) showed no change in grain structure, while it is observed that considerable reduction in dislocations and the formation of dislocation loops. This observation is consistent with gradual reduction of lifetime till 823 K. Notable observation is that while in CW Fe sample, lifetime reaches bulk value by 823 K (Fig-1(a)), AQ sample attains 118 ps, indicating these grain boundaries and dislocations loops are the positron trapping defects. 

For 900 K annealed sample (Fig-2(b)), the initial fine grain structure is retained and shows negligible amount of dislocations. But precipitates like contrast is observed uniformly throughout the sample, particularly dense along the grain boundaries. These contrast features seen in TEM at this temperature is due to Cr rich precipitates, similar to the observation of electron irradiated studies \cite{ezawa1994}. The observed increase in lifetime beyond 823 K (Fig-1a) can be understood in terms of these Cr precipitates. The interfaces between these precipitates and matrix can act as positron trapping sites, thereby increasing mean positron lifetime.

Towards understanding type of possible interfacial traps, positron lifetime calculations were carried out for defect free Fe-Cr [100] oriented interface as well as for Cr and Fe vacancies at interfacial region. Table I lists computed positron lifetimes. For defect free Fe-Cr interface, positron lifetime is observed to be 112 ps and positrons preferentially annihilate in Fe region. Based on this, it is understandable that defect free  coherent Cr precipitates cannot trap positrons, owing to lesser positron affinity for Cr than Fe \cite{Puska1989, haribabu2011}. But the presence of vacancies, either in Cr cluster or at the Fe/Cr interface region, are the possible trapping sites, according to our positron lifetime calculation results shown in Table I. The observed experimental lifetime at 900 K is rather large compared to bulk Fe-Cr defect free value(112 ps), indicating vacancy like trapping sites at the interface.  In this context these precipitates are considered to be incoherent with matrix.

The 973 K sample (Fig-2c) showed decrease in the number of precipitates and well developed coarse equiaxed grain structure, typical grain structure of ferrite, at the expense of fine grains, while a few fine grains retained their structure. Lifetime starts decreasing from 973 K and it is in good agreement with considerable decrease in number of precipitates, indicating the dissolution of Cr precipitates. Finally, TEM image at 1073 K showed no precipitates as they dissolve into matrix and a well developed grain structure with grain size of the order of few $\mu$m was observed which resulted in bulk life time of 108 ps. 

\begin{figure}
\begin{center}
\includegraphics*[width=7cm]{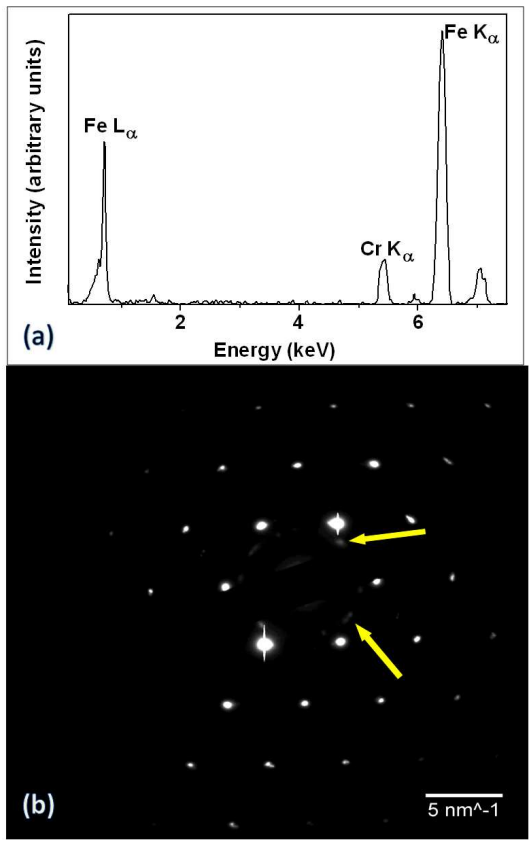}
\end{center}
\caption{\label{fig3} (a) EDS spectrum showing characteristic Fe and Cr peaks of a representative precipitate.  (b) SAD pattern of matrix along with embedded precipitates. The diffraction spots corresponding to additional phase are indicated with arrows.}
\end{figure}

We have further carried out EDS and SAD studies on 900 K annealed sample (corresponding to Fig 2(b)) using ATEM so as to confirm the nature of these precipitates. EDS spectrum obtained with a beam spot size of 7.5 nm (taken on a specific precipitate of size around 5 nm) is shown in fig 3(a) with characteristic Fe and Cr peaks.  The Cr concentration obtained is 12.6 $\pm$ 1.0 wt\%, while the Cr concentration away from the precipitate is found to be 8.9 $\pm$ 0.7 wt\%. As the size of precipitates is smaller than beam spot size and the film thickness is of the order of a few tens of nm, effective volume of precipitate is smaller compared to the matrix volume probed by EDS. Given the above considerations, the variation in the observed Cr concentration is reasonable to ascribe that these precipitates are Cr rich in nature. Fig 3(b) shows SAD pattern of matrix with embedded precipitates. Diffraction pattern showed an additional phase, other than BCC $\alpha$-Fe/Cr, with d-spacing value of 1.26 $\pm$ 0.12 $\rm{\AA}$. From JCPDS data, this value is consistent with both $\sigma$-phases of Fe-Cr as well as Fe-Cr carbides. In either case, the observation is in consistent with positron trapping at interfaces as they cannot be coherent with matrix. Formation of carbides is known due to supersaturation of carbon in BCC phase. In the present case, the concentration of carbon is well within the solubility limits and further, carbide precipitates are known to be stable and cannot dissolve around 973 K, as observed in the present case. Also, there are no such signatures of precipitates formation seen in HT sample to be consistent with the case of carbide formation. Thus, the current observations clearly point out that these are not carbide precipitates. Therefore, we attribute these precipitates to be Cr rich clusters occurring in Fe matrix. It is also interesting to note that the lower and upper limits of formation of $\sigma$-phase in FeCr system, as discussed in the recent review\cite{dubiel2011}, is in good agreement with the temperature range of Cr clustering in the present study (Fig 1(a)). However, exact phase identification is yet to be confirmed. 

Regarding HT sample(Fig 1(a)), it is expected that the as-prepared sample will have initial microstructure containing well developed grain structure without dislocations, as it is solution heat treated at 1073 K. The measured positron lifetime of 108 ps supports the consideration. So, it is understandable that there is no possibility of variation in lifetime until 823 K  due to dislocation annealing in HT sample. Further, lifetime in HT sample did not show any change during 823 K and 1023 K, as seen in AQ sample, indicating the absence of precipitation stage in HT sample. So, by comparing the features of HT and AQ samples, it is clear that initial microstructure, containing dislocations and fine grains, plays very important role in Cr precipitate in Fe-Cr system. 

\begin{figure}
\begin{center}
\includegraphics*[width=7.5cm]{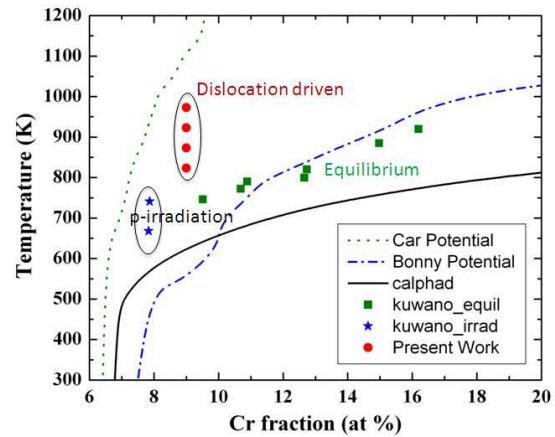}
\end{center}
\caption{\label{fig4} Fe rich side of phase diagram of Fe-Cr system. Calculated phase boundaries are shown as lines with Car potential \cite{car}, Bonny potential \cite{bonny2011} and CALPHAD \cite{calphad}. Filled squares represent experimental equillibrium phase boundary \cite{kuwano1985}, filled stars represent proton irradiation induced Cr clustering \cite{kuwano1988} and filled circles represent present study showing dislocarion driven Cr clustering between 823 and 973 K.}
\end{figure}

With regard to our observation of Cr precipitates in AQ samples, M. Matijasevic et al. \cite{matijasevic2008} found similar TEM contrast in neutron irradiated samples and ascribed them as radiation induced solute-defect clusters and their initial microstructure is similar to the microstructure observed in the present study. Incoherent precipitates formation at grain boundaries can be understood as fresh nucleation and growth. However, intra-granular incoherent precipitation is surprising. The mechanism of formation is believed to be dislocations acting as nucleating sites and by solute sweeping during dislocation recovery similar to proposal by Kesternich \cite{kesternich1985}. Accordingly, the precipitation of Cr is observed during the annealing out of pre-existing dislocations. According to the proposed mechanism moving dislocations sweep along solute atoms (if the solute atoms binding energy with dislocations is positive) till they react or two dislocations annihilate leaving locally enhanced concentration. The observed higher precipitate density along lath boundaries can be understood due annealing of dislocations at lath boundaries. This argument is further strengthened with evidence from absence of precipitation stage in the case of dislocation-free HT sample. Also, the temperature at which Cr precipitation occurs i.e. 873 K, coincides with the observed onset of dislocation annealing in pure Fe. While we see high concentration of precipitates along lath boundaries, earlier studies of precipitation by electron irradiation showed depletion of Cr from grain boundaries \cite{Wakai-1995, Heishichiro-1981}. These reports are understood as annihilation of irradiation induced dislocations within the matrix. Another interesting finding related to dependence of initial micrstructure is,  Wakai et al. \cite{Wakai-1995} used furnace cooled samples, expected to be coarse grained structure without dislocations, for irradiation and observed depletion of Cr at grain boundaries. Ezava et al. \cite{ezawa1994} used cold rolled 60 $\mu$m thick samples for irradiation and observed Cr enrichment at grain boundaries at 873 K. In the later case recrystallization of the deformed structure, to subgrain structure, is expected at this temperature. Heishichiro et al. \cite{Heishichiro-1981} also showed depletion of Cr at grain boundaries but not reported the initial state of the sample. Figure 4 shows the Fe-rich phase diagram taken from Bonny et. al. \cite{bonny2011} along with our experimental points corresponding to Cr clustering between 823 and 973 K. Though computed phase boundaries with Car \cite{car} and Bonny potentials \cite{bonny2011} were shown to be agreeing with CALPHAD computed phase boundary \cite{calphad} at low temperatures, they are  seen to deviate at high temperatures $>$ 500 K. For comparison, experimental equilibrium phase boundary data \cite{kuwano1985} as well as proton irradiation induced Cr clustering data \cite{kuwano1988} are shown in figure 3. It should also be noted that the critical temperature computed by  Bonny et. al. \cite{bonny2011} is 1200 K. It is clear from the present study that dislocations can also initiate Cr clustering, similar to that induced by irradiation \cite{kuwano1988}. The current results in terms of initial defect annealing and Cr precipitation at intermediate temperatures are relevant in prompting computational investigation including non-equilibrium nucleation sites like dislocations and lath boundaries as well as influence of dislocation dynamics on Cr clustering.  

\section{Conclusions}

Signatures are obtained for dislocation driven Cr precipitation in Fe-9Cr alloy during heat treatment from 823  to 1023 K from positron lifetime studies and these are supported by TEM observations. The precipitates are found to be chromium rich from Energy-Dispersive-X-ray analysis. The formation of Cr precipitates is found to be critically dependent on initial microstructure especially the presence of dislocations and fine grain structure. 

\section*{Ackowledgments}

One of the authors SHB acknowledge the research fellowship from the DAE, India. Authors acknowledge Shri. Varghese Anto Chirayath, MSG, IGCAR for useful discussions and comments, Dr. V. Chandramouli of CG, IGCAR and Mrs. C. Sudha of PMG, IGCAR for carbon analysis and EPMA respectively.

\end{document}